\documentclass[10pt, conference, compsocconf]{IEEEtran}

\usepackage{microtype}
\usepackage[ruled]{algorithm}
\usepackage{algpseudocode}
\usepackage{ioa_code}
\usepackage{amssymb,amsmath,multicol}

\usepackage{algorithm}
\usepackage{algpseudocode}
\usepackage{lipsum}

\usepackage{amsfonts}
\usepackage{epsfig}
\usepackage{subfigure}
\usepackage{calc}
\usepackage{color}
\usepackage[all]{xy}
\usepackage{bm}
\usepackage{enumerate}
\usepackage{pdfsync}
\usepackage{framed}
\usepackage{float}
\usepackage{cite}

\newcommand{\prf}[1]{{}}

\newtheorem{Def}{Definition}[section]

\newcommand{\sacode}[5]
{ \vspace{.06in} \hrule \vspace{.06in} 
 \noindent {\bf #1}: \\
 \footnotesize \noindent {\bf Signature:}\B \nobreak
 \normalsize \begin{quote} \nobreak #2 \end{quote}
 \footnotesize \noindent {\bf States:}\B \nobreak
 \begin{quote} \nobreak #3 \end{quote}
 \noindent {\bf Transitions:} \nobreak
 \vspace{-.2in} \nobreak
 \normalsize #4
 \vspace{-.06in} \hrule \vspace{.06in} 
}

\newcommand{\act}[1]{%
    \relax\ifmmode
        \mathord{\mathcode`\-="702D\sf #1\mathcode`\-="2200}%
    \else
        $\mathord{\mathcode`\-="702D\sf #1\mathcode`\-="2200}$%
    \fi
}

\newcommand{\tup}[1]{%
    \relax\ifmmode
      \langle #1 \rangle%
    \else
        $\langle$#1$\rangle$%
    \fi
}

\newcommand{\seq}[1]{%
    \relax\ifmmode
      \langle \! \langle #1 \rangle \! \rangle%
    \else
        $\langle \! \langle$ #1 $\rangle \! \rangle$%
    \fi
}

\newcommand{\B}{\vspace*{-\smallskipamount}}

\newcommand{\T}{\hspace*{1em}}
\newcommand{\TT}{\hspace*{2em}}

\newcommand{\ms}[1]{%
    \relax\ifmmode
        \mathord{\mathcode`\-="702D\it #1\mathcode`\-="2200}%
    \else
        {\it #1}%
    \fi
}

\newcommand{\lit}[1]{%
    \relax\ifmmode
        \mathord{\mathcode`\-="702D\sf #1\mathcode`\-="2200}%
    \else
        {\it #1}%
    \fi
}

\newcommand{\XDK}[1]{}
\newcommand{\remove}[1]{} 
\newcommand{\uselater}[1]{} 

\renewcommand{\setminus}{-}

\makeatletter
\def\mainlistofsymbols{
  \normalsize
  \vspace*{1.5 em}
  \@starttoc{los}
}

\def\partonelistofsymbols{
  \normalsize
  \vspace*{1.5 em}
  \@starttoc{p1los}
}

\def\parttwolistofsymbols{
  \normalsize
  \vspace*{1.5 em}
  \@starttoc{p2los}
}

\def\l@symbol#1#2{\addpenalty{-\@highpenalty} \vskip 4pt plus 2pt
{\@dottedtocline{0}{0em}{8em}{#1}{#2}}}
\makeatother

\newcommand{\newhiddensym}[2]{%
}

\newcommand{\algIOA}[2]{\ifmmode{\text{#1}_{#2}}\else{$\text{#1}_{#2}$}\fi}

\newcommand{\srvIOA}[2]{\algIOA{#1}{#2}}

\newcommand{\EX}{\ifmmode{\xi}\else{$\xi$}\fi}
\newcommand{\EXF}{\ifmmode{\phi}\else{$\phi$}\fi}

\newcommand{\inter}[1]{
	\ifmmode{\left(\bigcap_{\mathcal{Q}\in#1}\mathcal{Q}\right)}
	\else{$\left(\bigcap_{\mathcal{Q}\in#1}\mathcal{Q}\right)$}
	\fi
}

\newcommand{\op}{\pi}

\mathchardef\mhyphen="2D

\newcommand{\srvr}{s}

\newcommand{\vid}[1]{\ifmmode{\nu_{#1}}\else{$\nu_{#1}$}\fi}

\newcommand{\seen}{\ifmmode{seen}\else{$seen$}\fi}

\newcommand{\msgSet}{M}

\newcommand{\maxts}[1]{\ifmmode{maxTS_{#1}}\else{$maxTS_{#1}$}\fi}
\newcommand{\maxtag}[1]{\ifmmode{maxTag_{#1}}\else{$maxTag_{#1}$}\fi}
\newcommand{\maxpair}[1]{\ifmmode{maxMPair_{#1}}\else{$maxMPair_{#1}$}\fi}
\newcommand{\mintag}[1]{\ifmmode{minTag_{#1}}\else{$minTag_{#1}$}\fi}
\newcommand{\maxps}{\ifmmode{maxPS}\else{$maxPS$}\fi}
\newcommand{\conftg}[1]{\ifmmode{confirmed_{#1}}\else{$confirmed_{#1}$}\fi}
\newcommand{\maxconftag}{\ifmmode{\ms{maxCT}}\else{$maxCT$}\fi}

\newtheorem{theorem}{Theorem}[section]
\newtheorem{lemma}[theorem]{Lemma}

\newtheorem{definition}{Definition}

\newcommand{\readd}{{read}}
\newcommand{\GetR}{{ \it{read-get}}}
\newcommand{\GetRV}{{ \it{read-value}}}
\newcommand{\CompR}{{ \it{read-complete}}}
\newcommand{\GetW}{{ \it{write-get}}}
\newcommand{\PutW} {{ \it{write-put}}}

\newcommand{\GetRVResp}{{ \it{read-value}}}

\newcommand{\GetWResp}{{ \it{write-get}}}

\newcommand{\SODA}{{ \act{SODA}}}

\newcommand{\mdisperse}{ {\em message-disperse}}

\newcommand{\writeGetTag}{{\sc write-get}}
\newcommand{\readValueTag}{{\sc read-value}}
\newcommand{\readGetTag}{{\sc read-get}}
\newcommand{\readCompleteTag}{{\sc read-complete}}
\newcommand{\readDisperseTag}{{\sc read-disperse}}

\newcommand{\mdmetasend}{{\act{md-meta-send}}}

\newcommand{\mdvaluesend}{{\act{md-value-send}}}

\newcommand{\mdmetaprim}{{\sc md-meta}}
\newcommand{\mdvalueprim}{{\sc md-value}}

\newcommand{\mdvaluesender}{{\sc md-value-sender}}
\newcommand{\mdvalueserver}{{\sc md-value-server}}

\begin{document}

\title{Storage-Optimized Data-Atomic Algorithms for Handling Erasures
and Errors in Distributed Storage Systems}

\author{\IEEEauthorblockN{Kishori M. Konwar$^{1}$, N. Prakash$^{1}$, Erez Kantor$^{2}$,  Nancy Lynch$^{1}$, Muriel M{\'{e}}dard$^{1}$, and Alexander A. Schwarzmann$^{3}$}
\IEEEauthorblockA{$^{1}$Department of EECS, MIT, MA, USA, $^{2}$Department of Computer Science, NEU, MA, USA,  \\ 
$^{3}$Department of CSE, UConn., Storrs, CT, USA}
}
\maketitle

\begin{abstract}
	
	\emph{Erasure codes} are increasingly being studied in the context of implementing atomic memory objects in large scale asynchronous distributed storage systems. When compared with the traditional replication based schemes, erasure codes have the potential of significantly lowering storage and communication costs while simultaneously guaranteeing the desired resiliency levels.  In this work, we propose the Storage-Optimized Data-Atomic (SODA) algorithm for implementing atomic memory objects in the multi-writer multi-reader setting. SODA uses Maximum Distance Separable (MDS) codes, and is specifically designed  to optimize the total storage cost for a given fault-tolerance requirement. For tolerating $f$ server crashes in an $n$-server system, SODA uses an $[n, k]$ MDS code with $k=n-f$, and incurs a  total storage  cost of $\frac{n}{n-f}$. SODA is designed under the assumption of reliable point-to-point communication channels. The communication cost of a write and a read operation are respectively given by $O(f^2)$ and   $\frac{n}{n-f}(\delta_w+1)$, where $\delta_w$ denotes the number of writes that are concurrent with the particular read. In comparison with the recent CASGC algorithm~\cite{CadambeLMM14}, which also uses MDS codes, SODA offers lower storage cost while pays more on the communication cost.
	
	We also present a modification of SODA, called SODA$_{\text{err}}$,  to handle the case where some of the servers can return  erroneous coded elements during a read operation. Specifically, in order to tolerate $f$ server failures and $e$ error-prone coded elements, the SODA$_{\text{err}}$ algorithm uses an $[n, k]$ MDS code such that $k=n-2e-f$. SODA$_{\text{err}}$ also guarantees liveness and atomicity, while maintaining an optimized total storage cost of $\frac{n}{n-f-2e}$.
	
	
\end{abstract}


%
\IEEEpeerreviewmaketitle

\section{Introduction} \label{sec:intro}

The demand for efficient and reliable large-scale distributed storage systems (DSSs) has grown at an unprecedented scale in the recent years. DSSs that store massive data sets across several hundreds of  servers are increasingly being used for both industrial and scientific applications, ranging from  sequencing genomic  data to those used for e-commerce. Several applications demand concurrent and consistent access to the stored data by multiple writers and readers. The consistency model  we adopt is {\em atomicity}. Atomic consistency gives the users of the data service the impression that the various concurrent read and write operations happen sequentially. Also, ability to withstand failures and network delays  are essential features of any robust DSS.

The traditional solution for emulating an atomic fault-tolerant shared storage system  involves  replication of data across the servers. Popular replication-based algorithms appear in the work by Attiya, Bar-Noy and Dolev~\cite{ABD96} (we refer to this as the ABD algorithm) and also in the work by Fan and Lynch~\cite{FL03} (which is referred to as the LDR algorithm). Replication based strategies incur high storage costs; for example, to store a {\em value} (an abstraction of a data file) of size $1$ TB across a $100$ server system, the ABD algorithm replicates the value in all the $100$ servers,  which blows up the worst-case {\em storage cost} to $100$ TB. Additionally, every write or read operation has a worst-case {\em communication cost} of $100$ TB. The communication cost, or simply the cost, associated with a read or write operation is the amount of total data in bytes that gets transmitted in the various messages sent as part of the operation. Since the focus in this paper is on  large data objects, the storage and communication costs include only the total sizes of stable storage and messages dedicated to the data itself. Ephemeral storage and the cost of control communication is assumed to be negligible. Under this assumption, we further {\em normalize} both the storage and communication costs with respect to the size of the value, say $v$, that is written, i.e.,  we simply assume that the size of $v$ is $1$ unit (instead of $1$ TB), and say that the worst-case storage or read or write cost of the ABD algorithm is $n$ units, for a system consisting of $n$ servers.  

Erasure codes provide an alternative way to emulate fault-tolerant shared atomic storage. In comparison with replication, algorithms based on erasure codes significantly reduce both the storage and communication costs of the implementation. An $[n, k]$ erasure code splits the value $v$ of size $1$ unit into $k$ elements, each of size $\frac{1}{k}$ units, creates $n$ {\em coded elements}, and stores one coded element per server. The size of each coded element is also $\frac{1}{k}$ units, and thus the total storage cost across the $n$ servers is $\frac{n}{k}$ units. For example, if we use an $[n = 100, k = 50]$ MDS code, the storage cost is simply  $2$ TB, which is almost two orders of magnitude lower than the storage in the case of ABD. A class of erasure codes known as Maximum Distance Separable (MDS) codes have the property that value $v$ can be reconstructed from any $k$ out of these $n$ coded elements. In systems that are centralized and synchronous, the parameter $k$ is simply chosen as $n-f$, where $f$ denotes the number of server crash failures that need to be tolerated. In this case, the read cost, write cost and total storage cost can all be simultaneously optimized. The usage of MDS codes to emulate atomic shared storage in decentralized, asynchronous settings is way more challenging, and often results in additional communication or storage costs for a given level of fault tolerance, when compared to the synchronous setting. Even then, as has been shown in the past~\cite{CadambeLMM14}, \cite{DGL08}, significant gains over replication-based strategies can still be achieved while using erasure codes.  In \cite{CadambeLMM14} and \cite{DGL08} contain algorithms based on MDS codes for emulating fault-tolerant shared atomic storage, and offer different trade-offs between storage and communication costs. 

\subsection{Our Contributions} \label{sec:contrib}

In this work we propose the {\em Storage-Optimized Data-Atomic} (SODA) algorithm for implementing atomic memory objects. SODA uses $[n, k]$ MDS codes, and is specifically designed to optimize the total storage cost for a given fault tolerance level. We also present a modification of SODA, called SODA$_{\text{err}}$,  in order to handle the case where some of the non-faulty servers can return erroneous coded elements during a read operation. A summary of the algorithms and their features are provided below:
\vspace{0.05in}
\paragraph{The SODA Algorithm} SODA assumes reliable point-to-point communication channels between any two processes - the collection of all readers, writers and servers - in the system. In a system consisting of $n$ servers, for tolerating $f, 1 \leq f  \leq \frac{n-1}{2}$ server crashes, SODA uses an $[n, k]$ MDS code with $k = n-f$. Each server at any point during the execution of the algorithm  stores at most one coded element, and thus, SODA has a worst-case total storage cost of $\frac{n}{n-f}$.
We prove the {\em liveness} and {\em atomicity} properties of SODA in the multi-writer multi-reader (MWMR) setting, for executions in which at most $f$ servers crash. Any number of writer or reader processes may fail during the execution. 

We construct a {\em message-disperse} primitive and use it in  the write and read operations in SODA. The primitive is used by a process $p$ to disperse a message $m$ to all the non-faulty servers. The message $m$ can be either {\em meta-data} alone or one that involves the value $v$ along with a tag (where the tag is used to identify the version associated with the value); slightly differing implementations are used in the two cases. Meta-data refers to data such as ids, tags etc. which are used by various operations for book-keeping. In situations where $m$ consists only of meta-data, the primitive ensures that if a server $s \in S$ receives $m$, then the same message $m$ is sent to every server $s' \in S$ by some process in the set $\{p\} \cup S$. Thus if $s'$ is non-faulty, it eventually receives $m$ since the point-to-point channels are assumed reliable.  During write operations, the writer uses the {\em message-disperse} primitive where $m$ is the value $v$ to be written. In this case, the primitive ensures that every non-faulty server receives the coded element that is targeted for local storage in that server. The primitive can tolerate up to $f$ server failures and also the failure of the process $p$. The idea here is to ensure that the uncoded value $v$ is sent to $f+1$ servers, so that at least one non-faulty server receives $v$. This non-faulty server further computes and sends the corresponding coded elements to the remaining $n-f$ servers.  We show that the communication cost for a write operation, implemented on top of the {\em message-disperse} primitive, is upper bounded by $5f^2$.

The read operations in SODA use a reader-registration and relaying technique similar to the one used in \cite{cachin}, where the authors discuss the use of erasure codes for Byzantine fault tolerance. For successful decoding, a reader must collect $k$ coded elements corresponding to one particular tag. The reader registers itself with all non-faulty servers, and these servers send their respective (locally stored) coded elements back to the reader. Further, each non-faulty server also sends to the reader the coded elements it receives as part of concurrent write operations. Such relaying, by the servers, is continued until the reader sends a message acknowledging read completion.  SODA uses a  server-to-server communication mechanism to handle the case where a reader might fail after invoking a read operation. This internal communication mechanism  exchanges only metadata and ensures that no non-faulty server relays coded elements forever to any reader. No such mechanism is used in \cite{cachin} to handle the case of a failed reader. The read cost of SODA is given by  $\frac{n}{n-f}(\delta_w+1)$, where $\delta_w$ denotes the number of writes that are concurrent with the particular read. Since  $\delta_w$ might vary across different reads, the cost also varies across various reads, and hence we say that the read cost is {\em elastic}. The parameter $\delta_w$  appears only as part of the analysis; its knowledge is not necessary to ensure liveness or atomicity. 

We also carry out a latency analysis of successful write/read operations in SODA. The analysis assumes that latency arises only from the time taken for message delivery, and that computations at processes are fast. Under the assumption that the delivery time of any message is upper bounded by $\Delta$ time units, we show that every successful write and read operation completes in  $5\Delta$ and $6\Delta$ time units, respectively. The read time in this model of latency analysis turns out to be independent of the number of concurrent writers in the system.
\vspace{0.05in}
\paragraph{The SODA$_{\text{err}}$ Algorithm} The SODA$_{\text{err}}$ algorithm is designed to handle the additional case where some of the servers can return erroneous coded elements during a read operation.  The added feature of the algorithm is useful in large scale DDSs, where  commodity hard disks are often used to achieve scalability of storage at low costs. In such systems, a coded element accessed by the server from its local hard-disk can be  erroneous, i.e., the server obtains an arbitrary valued element instead of what was expected; however the server is not aware of the error when it sends this element back to the reader. The SODA$_{\text{err}}$ algorithm provides a framework for tackling local disk read errors via the overall erasure code across the various servers, without the need for expensive error protection mechanisms locally at each server. Specifically, in order to tolerate $f$ server failures (like in SODA) and $e$ error-prone coded elements, SODA$_{\text{err}}$ uses an $[n, k]$ MDS code such that $n - k = 2e + f$. We assume that no error occurs either in meta data or in temporary variables, since these are typically stored in volatile memory instead of local hard disk. SODA$_{\text{err}}$ also guarantees liveness and atomicity in the MWMR setting, while maintaining an optimized total storage cost of $\frac{n}{n-f-2e}$.  The write cost is upper bounded by $5f^2$, and the read cost is given by $\frac{n}{n-f-2e}(\delta_w + 1)$.

\subsection{Comparison with Other Algorithms, and Related Work}

We now compare SODA with the algorithms in \cite{CadambeLMM14} and \cite{DGL08}, which are also based on erasure codes for emulating fault-tolerant atomic memory objects. In \cite{CadambeLMM14}, the authors provide two algorithms - CAS and CASGC - based on $[n, k]$ MDS codes, and these are primarily motivated with a goal of reducing the communication costs. Both algorithms tolerate up to $f = \frac{n-k}{2}$ server crashes, and incur a communication cost  (per read or write) of $\frac{n}{n-2f}$. The CAS algorithm is a precursor to CASGC, and its storage cost is not optimized. In CASGC, each server stores coded elements (of size $\frac{1}{k}$) for up to $\delta+1$ different versions of the value $v$, where $\delta$ is an upper bound on the number of writes  that are concurrent with a read. A garbage collection mechanism, which removes all the older versions, is used to reduce the storage cost.  The worst-case total storage cost of CASGC is shown to be   $\frac{n}{n-2f}(\delta+1)$. Liveness and atomicity of CASGC are proved under the assumption that the number of writes concurrent with a read never exceeds $\delta$. In comparison, SODA is designed to optimize the storage cost rather than communication cost. We now note the following important differences between CASGC and SODA. $(i)$ In SODA, we use the parameter $\delta_w$, which indicates the number of writes concurrent with a read, to bound the read cost. However, neither liveness nor atomicity of SODA depends on the knowledge of $\delta_w$; the parameter  appears only in the analysis and not in the algorithm. $(ii)$ While the effect of the parameter $\delta$ in CASGC is rather {\em rigid}, the effect of $\delta_w$ in SODA is elastic. In CASGC, any time after $\delta+1$ successful writes occur during an execution, the total storage cost remains fixed at $\frac{n}{n-2f}(\delta+1)$, irrespective of the actual number of concurrent writes during a read. $(iii)$ For a given $[n, k]$ MDS code, CASGC tolerates only up to $f = \frac{n-k}{2}$ failures, whereas SODA tolerates up to $f =  n - k$ failures. A comparison of the performance numbers at $f_{\max} = \left\lfloor\frac{n-1}{2}\right\rfloor$  is show in Table \ref{table:summ}. Note that $f_{max}$ is the maximum number of failures for which either of these algorithms can be designed. Also note that $f_{max}$ denotes the maximum  of failures that can be tolerated by the ABD algorithm, as well.
\begin{table}[!h]
	\centering
	\begin{tabular}{c c c c}
		\hline
		Algorithm & \emph{Write} Cost & \emph{Read} Cost   & Total storage cost \\ [0.5ex] 
		\hline
		ABD & $n$ & $n$  & $n$  \\
		CASGC &$\frac{n}{2}$ & $\frac{n}{2}$ &  $\frac{n}{2}(\delta+1) $ \\
		SODA & $O(n^2)$ & $\leq 2(\delta_w+1)$  &  $\leq 2$ \\
		\hline
	\end{tabular}
	\caption{Performance comparison of ABD, CASGC and SODA, for $f = f_{\max} = \frac{n}{2}-1$. We assume $n$ to be an even number.} \label{table:summ}
\vspace{-0.4in}	
\end{table}

In \cite{DGL08}, the authors present the ORCAS-A and ORCAS-B algorithms for asynchronous crash-recovery models. In this model, a server is allowed to undergo a temporary failure such that when it returns to normal operation, contents of temporary storage (like memory) are lost while those of permanent storage are not. Only the contents of permanent storage count towards the total storage cost.  
Furthermore they do not assume reliable point-to-point channels.  
The ORCAS-A algorithm  offers better storage cost than ORCAS-B when the number of concurrent writers
is small. Like SODA, in ORCAS-B also coded elements corresponding to multiple versions are sent by a writer to reader, until the read completes. However, unlike in SODA, a failed reader might cause servers to keep sending coded elements indefinitely. We do not make an explicit comparison of storage and communication costs between SODA and ORCAS because of the difference in the models.

In \cite{kedar_bounds}, the authors consider algorithms that use erasure codes for emulating {\em regular} registers. Regularity~\cite{regular_lamport}, \cite{shao} is a weaker consistency notion than atomicity. Distributed storage systems based on erasure codes, and requiring concurrency/consistency  are also considered in \cite{aguilera}. Applications of erasure codes to Byzantine fault tolerant DSSs are discussed in \cite{cachin}, \cite{dobre}, \cite{hendricks}. RAMBO~\cite{LS02} and  DynaStore~\cite{AKMS11} are implementations of MWMR atomic memory objects in dynamic DSSs, where servers can enter or leave the system. 
\vspace{0.05in}
\paragraph{Document Structure} Models and definitions appear in Section \ref{sec:models}. Implementation and properties of the {\em message-disperse} primitives are discussed in Section \ref{sec:md}.   Description and analysis of the SODA algorithm are in Sections  \ref{sec:algo} and \ref{sec:analysis}, respectively. SODA$_{\text{err}}$ algorithm is presented in Section \ref{sec:soda_err}.  Section \ref{sec:conclusions} concludes. Due to space constraints, proofs are omitted. 

\section{Models and definitions} \label{sec:models}
In this section, we describe the models of computation,  explain the concepts of atomicity, erasure codes, and the performance metrics used in the paper.
\vspace{0.05in}
\paragraph{Asynchrony and Crash Failures}
We consider a distributed system consisting of \emph{asynchronous} processes of three types: a set of \emph{readers} and  \emph{writers}, called clients, and a set of $n$ \emph{servers}.  Each of these processes is associated with a unique identifier, and we denote the sets of IDs of the readers, writers and servers as ${\mathcal R}$, ${\mathcal W}$ and  ${\mathcal S}$, respectively. The  set of IDs forms  a  totally ordered set. The reader and writer processes initiate {\em read} and {\em write}  operations, respectively,  and communicate with the servers using messages. Also, any client initiates a new operation only after the previous operations, if any,  at the same client has completed. We refer to this as the {\em well-formedness} property of an execution. All processes run local computations until completion or crash failure.  Any of the number of clients  can fail. We assume up to $f$, such that,  $f \leq \frac{n-1}{2}$,  servers (out of the total $n$)  may crash during any execution.

We assume that every client is connected to every server through a reliable communication
link. This means that as long as the destination process is non-faulty, any message sent on  the link is guaranteed to eventually reach the destination process.  The model allows the sender
process to fail after placing the message in the channel; message-delivery depends
only on whether the destination is non-faulty. We also assume reliable connectivity between every pair of servers in the system. We do not make any assumption regarding relative order of message delivery in the same channel.
  
\vspace{0.05in}
\paragraph{Liveness}
By liveness, we mean that during any well-formed execution of the algorithm,  any read or write operation initiated by non-faulty reader or writer completes,  despite the crash failure of any other clients and up to $f$ server crashes.

\vspace{0.05in}
\paragraph{Atomicity}
A shared atomic memory can be emulated by composing individual atomic objects.
Therefore, we aim to implement only one atomic read/write memory object, say $x$, on a  
set of  servers.
The object value $v$ comes from some set $V$; initially 	$v$ is set to a distinguished value $v_0$ ($\in V$). Reader $r$ requests a read operation on  object $x$.
Similarly, a write operation is requested by a writer $w$. Each operation at a non-faulty client begins with an \emph{invocation step} and terminates with a  \emph{response step}.
An operation $\op$ is \emph{incomplete} in an execution when the invocation step of $\op$ does not have the associated response step; otherwise we say that $\op$ is \emph{complete}. 
In an execution, we say that an operation (read or write) $\op_1$ {\em precedes} another operation $\op_2$,  if the response step for $\pi_1$ precedes the invocation step of $\pi_2$.  Two operations 
are {\em concurrent} if neither precedes the other. The following lemma is a restatement of the sufficiency condition for atomicity  presented in~\cite{Lynch1996}.

\begin{lemma} \label{lem:atom}
	For any execution of a memory service, if all the invoked read and the 
	write operations  are complete, then the 
	operations can be partially ordered by an ordering $\prec$, so that the 
	following properties are satisfied:
	\begin{itemize}
		\item [\em P1.] The partial order ($\prec)$ is consistent with the 
		external order of invocation and responses, i.e., there are no 
		operations $\pi_1$ and $\pi_2$, 
		such that $\pi_1$ completes before $\pi_2$ starts, 
		yet $\pi_2 \prec \pi_1$.
		\item[\em P2.] All operations are totally 
		ordered with respect to the write operations, 
		i.e., if $\pi_1$ is a write operation and $\pi_2$ is any other operation then 
		either $\pi_1 \prec \pi_2$ or $\pi_2 \prec \pi_1$.
		\item[\em P3.] Every read operation ordered after any writes returns
		the value of the last write preceding it (with respect to $\prec$), and if no preceding 
		writes is ordered before it then it returns the initial value
		of the object.
	\end{itemize}
\end{lemma}	
\vspace{0.05in}
\paragraph{Erasure coding} We use $[n, k]$  linear MDS codes~\cite{verapless_book} to encode and store the value $v$ among the $n$ servers. An $[n, k, d]$ linear code $\mathcal{C}$ over a finite field $\mathbb{F}_q$ (containing $q$ elements) is a $k$-dimensional subspace of the vector space $\mathbb{F}_q^n$. The parameter $d$ is known as the minimum distance of the code $\mathcal{C}$ and denotes the minimum Hamming weight of any non-zero vector in $\mathcal{C}$. The well known Singleton bound~\cite{singleton} states that $d \leq n - k + 1$. Codes that achieve equality  in  this bound  are known as MDS codes, such codes are   known to exist for any $(n, k )$ pair such that $k \leq n$, e.g.,  Reed-Solomon codes~\cite{rscodes}.

We use functions  $\Phi$ and $\Phi^{-1}$ to denote the encoder and decoder associated with the code $\mathcal{C}$.  For encoding, $v$ is divided into $k$ elements $v_1, v_2, \ldots v_k$ with each element having a size $\frac{1}{k}$. As mentioned in Section \ref{sec:intro}, we assume that the value $v$ is of size $1$ unit. The encoder takes the $k$ elements as input and produces $n$ coded elements $c_1, c_2, \ldots, c_n$ as output, i.e., $[c_1, \ldots, c_n] = \Phi([v_1, \ldots, v_k])$. For ease of notation, we will simply write $\Phi(v)$ to mean  $[c_1, \ldots, c_n]$. The vector $[c_1, \ldots, c_n]$ is often referred to as the codeword corresponding to the value $v$. Each coded element $c_i$ also has a size $\frac{1}{k}$. In our scheme we store one coded element per server. We use $\Phi_i$ to denote the projection of $\Phi$ on to the $i^{\text{th}}$ output component, i.e., $c_i = \Phi_i(v)$. Wlog, we associate the coded element $c_i$ with server $i$, $1 \leq i \leq n$. 

A code with minimum distance $d$ can tolerate up to $d-1$ erasures among the $n$ coded elements. Since we wan tolerate up to $f$ server failures while using MDS codes, we pick the dimension $k$ of the MDS code as $k = n -f$. Since we store one coded element per server,  by using an $[n, n-f]$ MDS code, we get the property that the original value $v$ can be recovered given the contents of any $n-f$ servers.  If $C = \{c_i, i \in \mathcal{I}\}$ denotes any multiset of $k$ coded elements for some $\mathcal{I} \subset [n], |\mathcal{I}| = k$, we write $v = \Phi^{-1}(C)$ to indicate that $v$ is decodable from $C$. We implicitly assume that the process that is invoking the decoder is aware of the index set $\mathcal{I}$ corresponding to the $k$ coded elements. 
\paragraph{Storage and Communication Cost} \label{sec:sc} We define the (worst-case) total storage cost 
as the size of the data stored across all servers, at any point of the execution of the algorithm.
As mentioned in Section \ref{sec:intro}, the storage cost is normalized with respect to the size of the value $v$, which is equivalent to computing the storage cost under the assumption that $v$ has size $1$ unit. We assume metadata, such as version number, process ID,  used by various operations is of  negligible size and is hence ignored in the calculation of storage or communication cost. The communication cost associated with a read or write operation is the size of the total data that gets transmitted in the messages sent as  part of the operation. As with storage cost, we ignore the communication cost associated with metadata transmissions.
\section{The {\em message-disperse} primitives} \label{sec:md}
Now we discuss  the {\mdisperse}  services that are used to disseminate messages in {\SODA}. They have the property that if a message $m$ is delivered to any server in ${\mathcal S}$, then the same
message (or a derived message) is  eventually delivered at every non-faulty server in ${\mathcal S}$. The
services are implemented on  top of point-to-point reliable channels.
The services are provided in terms of $(i)$ the \emph{\mdmetaprim}~primitive, used for the metadata delivery, and  $(ii)$ the \emph{\mdvalueprim} primitive, used for delivering the coded elements for the  values.  The $\text{\mdmetaprim}$ (or $\text{\mdvalueprim}$) primitive is invoked by the send-event  \emph{md-meta-send} (or \emph{md-value-send}) at some process $p$, and results in delivery-event \emph{md-meta-deliver} (or \emph{md-value-deliver}) at any non-faulty process $s \in \mathcal{S}$. In order to reason about the properties of the protocols, we require precise descriptions of the flow of messages among the client, server and communication channel processes. Therefore, we specify their implementations using the language of IO Automata (see Lynch~\cite{Lynch1996}). Due to space constraints, only the $\text{\mdvalueprim}$ primitive is discussed in detail. The $\text{\mdmetaprim}$ primitive differs from the $\text{\mdvalueprim}$ primitive only in a minor way, and the difference alone will be discussed. 

\subsection{\text{\mdvalueprim}~primitive} 
The \mdvalueprim~primitive is to be used in {\SODA} to deliver the coded elements and the associated tags, which are  unique version identifiers for the object value,  to every non-faulty server.  Below we first define  the primitive, and its desired consistency properties. Subsequently, we present the implementation of the primitive. 
\vspace{0.05in}

\begin{definition}
\emph{\bf \mdvalueprim~primitive} sends message containing tag $t$ and value $v$  from a sender process  $p \in \mathcal{S}$ to the set of server processes in $\mathcal{S}$, such that  each non-faulty process in $\mathcal{S}$  delivers its corresponding coded elements. The following events define the primitive, to an external user\footnote{The $\act{md-value-send}$ and $\act{md-value-deliver}$ are the events that are used by the SODA algorithm.}: $(i)$ $\act{md-value-send}(t,v)_p$: an invocation event, at a writer $p \in \mathcal{W}$, that submits the version $t$ and the value $v$ for delivery of the coded elements, and $(ii)$ $\act{md-value-deliver}(t, c_p)_p$: an output event, at server $p \in \mathcal{S}$, that delivers the coded element $c_p = \Phi_p(v)$  to  the server $p$.
\end{definition}

Following are the consistency properties that we expect from an implementation (also called as protocol) of the primitive, under the assumption that all executions are well-formed.

\begin{definition}
\emph{\bf Consistency-Properties} $(i)$ \emph{validity}: if event $\act{md-value-deliver}(t,c_s)_s$ takes place at some server $s\in \mathcal{S}$,	then it is preceded by the event $\mdvaluesend(t, v)_w$ at a writer $w$, where $t \in \mathcal{T}$ and $c_s  = \Phi_s(v)$; and
$(ii)$ \emph{uniformity}: if event $\act{md-value-deliver}(t, c_s)_s$ takes place at some server $s \in \mathcal{S}$, and as long as the number of server crashes during the execution is at most $f$, then the event $\act{md-value-deliver}(t, c_{s'})_{s'}$ occurs at every  non-faulty process 	$s' \in \mathcal{S}$, where $c_{s'} = \Phi_{s'}(v)$. 
\end{definition}

We note that the uniformity property must hold even if the writer $w$ itself crashes after the invocation of the event $\mdvaluesend(t, v)_w$. 
\vspace{0.05in}

{\bf Implementation:} The IO Automata specifications of a sender, $\text{\mdvaluesender}_p,  
p \in \mathcal{W}$,  and the receiving servers, $\text{\mdvalueserver}_s$, $s \in \mathcal{S}$, for the $\text{\mdvalueprim}$~protocol are  given in Figs.~\ref{ioa:md-value-sender}~and~\ref{ioa:md-value-server}, respectively. The overall protocol is obtained by composing the above two automata and the  underlying automata for the point-to-point reliable channels (see Lynch~\cite{Lynch1996}). We first describe the data types, state variables and transitions that appear in the protocol, and then present a description of the protocol.

\vspace{0.05in}
\emph{Data Types and State Variables:}
In the  IO Automata specification of  $\text{\mdvalueprim}$, 
 for any value $v \in V$ the coded element corresponding to $s \in \mathcal{S}$ is denoted as $c_s \equiv \Phi_s(v)$.   
$M_{ID} \equiv {\mathcal S} \times \mathbb{N}$ is the set of unique message identifiers. 
Each message is one of two types: $\act{\sc TYPES} = \{``full", ``coded"\}$.
In  $\text{\mdvaluesender}_p$  boolean state variables $failed$ and 
$active$  are initially  $false$.
The  state variable, 
$mCount$, keeps track of  the number of times  $\text{md-value-send}(*)_p$  has been invoked at sender process $p$, and initially this is $0$.
The variable $send\_buff$ is a  FIFO queue  with elements of the form   $(M_{ID} \times  (\mathcal{T} \times V) \times \act{\sc TYPES}) \times \mathcal{S}$, and initially this is empty.
State variable $mID \in M_{ID}$  holds a unique message identifier corresponding to an invocation of the protocol, initially $(0, p)$.
Variable $currMsg$  holds the message that is being sent, initially  $\bot$.
In an automaton $\text{\mdvalueserver}_s$ we have the following state variables. 
The state variable $failed$  is initially set to $false$.
The variable $status$ is a  map  from keys in $M_{ID}$ to a value in $\{\act{ready}, \act{sending}, \act{delivered}\}$.  
The variable $content$ is a  map  from keys in $M_{ID}$ to a message in $\mathbb{F}_q $, initially $\bot$.  
The variable $outQueue$ is a  FIFO queue with elements of the form   $ M_{ID} \times  (V\cup \mathbb{F}_q ) \times \act{\sc TYPES}$, initially empty.
\emph{Transitions:}  In 
$\text{\mdvaluesender}_p$
the input action $\text{md-value-send}(t, v)_p$ invokes the protocol with tag $t$ and value $v$, 
and the output transition 
$\text{md-value-send-ack}(t, v)_p$ occurs when  all the messages with $t$ and $v$ are sent. The  action $\text{send}(*)_p$ adds messages to the channels.
Automaton $\text{\mdvalueserver}_s$ has two input actions $\act{recv}(*)_{*,s}$ corresponding to the ``full'' and ``coded'' types for receiving the values and coded elements, respectively, and 
the action $\text{send}(*)_s$ sends message to other servers through the channels. The output action $\text{md-value-deliver}(t,c)_s$
delivers the tag $t$ and coded element $c$ corresponding to server $s$.
{\em Explanation of the Protocol:} The basic idea of the $\text{\mdvalueprim}$ implementation is as follows: the sender $p \in {\mathcal W}$ invokes input action $\text{md-value-send}(t, v)_p$ at the automaton $\text{\mdvaluesender}_p$. The tag $t = \text{``full''}$ and value $v$ are sent to the set of first $f+1$ servers $D = \{s_1, s_2, \cdots, s_{f+1} \}$ among the set of all servers. Recall that in our model, we assume an ordering of the $n$ servers in the system, and hence it makes sense to talk about the first $f+1$ servers. Further, the message $m = (t, v)$ is sent to the servers respecting the ordering of the servers, i.e., $p$ sends $m$ to $s_i$ before sending to $s_{i+1}, 1 \leq i \leq f$.  

Let us next explain $\text{\mdvalueserver}_s, s \in S$. In this, let us first consider the case when $s = s_i \in  D = \{s_i, 1 \leq i \leq f+1\}$. In this case, the server $s_i$ upon receiving $m$ for the first time, sends $m$ to every process in the set  $\{s_{i+1}, s_{i+2}, \ldots, s_{f+1}\}$. Once again the message is sent to these $f+1-i$ servers respecting the ordering of the servers. As a second step, the server $s_i$, for every  server $s' \in \mathcal{S} \setminus D$, computes the coded element $c_{s'} = \Phi_{s'}(v)$ and sends the message $(t = \text{``coded''}, c_{s'} = \Phi_{s'}(v))$ to the server $s'$. Finally, the server $s_i$ computes its own coded element $c_{s_i} = \Phi_{s_i}(v)$ and delivers it locally via the output action $\text{md-value-deliver}(t, c_{s_i})_{s_i}$. Let us next consider the case when $s \in S - D$. In this case, the server $s$ simply delivers the received coded-element $c_{s}$  via the output action $\text{md-value-deliver}(t, c_{s})_{s}$. 
Next, we  claim the properties of the protocol.  

\begin{theorem}
	\label{thm:multicast_property_value}
	Any well-formed execution of the $\text{\mdvalueprim}$ protocol satisfies the consistency properties.
\end{theorem}
Next  theorem says that once a message is delivered via the primitive, all the associated messages get automatically removed from the system, i.e., there is no bloating-up of state variables.
\begin{theorem}\label{thm:no_garbage}
	Consider a well-formed execution $\beta$ of the  $\text{\mdvalueprim}$~protocol such that the  event  $\text{md-value-send}(t, v)_p$ 
	appears in $\beta$. Then, for any $s \in \mathcal{S}$  there exists a state $\sigma$ in $\beta$ after the event $\text{md-value-send}(t, v)_p$ 	such that in the automatons $\text{\mdvaluesender}_p$ and  $\text{\mdvalueserver}_s$ for every $s \in \mathcal{S}$,  the following is true : $(i)$ either $failed_s$ is $true$ or $(ii)$ in any state in $\beta$ following  $\sigma$,  none of the state variables in automatons contains  $v$ or  any of the coded elements $c_{s'}$ for $s' \in \mathcal{S}$.
\end{theorem}

\begin{algorithm*}[!ht]
	\caption{$\srvIOA{\text{\mdvaluesender}}{p}$ Automaton: Signature, State and Transitions at sender $p \in \mathcal{W}$.} \label{ioa:md-value-sender}
	\begin{algorithmic}[2]
	\begin{multicols}{3}
	{\scriptsize
	

\remove{
    	\State {\bf Data Types:}
                \State \T $M \subset M_{meta} \times (\mathbb{F}_q^k \cup \{ \bot\}) \times 
( \mathbb{F}_q  \cup \{ \bot\})$, 
                \State \T $M_{ID} \subseteq \mathcal{P} \times \mathbb{N}$,
     alphabet of  message identifiers

%
    \Statex
}
	\Part{Signature}{ \label{line:valueserver-sig}
		\State {\bf Input:}
		\State \T  $\act{md-value-send}(t, v)_{p}$, $t \in \mathcal{T}$, $v \in V$
				\State {\bf Internal:}
		\State \T{ $\act{fail}_{s}$}
		\State {\bf Output:}
		\State \T  $\act{send}((mID,  (t, v),  ``full"))_{p, s}$, 
		\Statex \TT\TT\T $mID \in M_{ID}$,  $t \in \mathcal{T}$, $v \in V$
		\State \T $ \act{md-value-send-ack}(t)_{p}$, $t \in \mathcal{T}$

	}\EndPart 
	\Statex\Statex

	\Part{State}{ \label{line:abdserver-state1}
		\State \T$failed$, a Boolean, initially $false$
               \State \T$active$, a Boolean, initially $false$
                 \State \T $mCount$, an integer, initially $0$
		 \State \T$send\_buff$, a queue,  
 initially $\emptyset$

                    \State \T $mID \in \mathbb{N} \times \mathcal{S}$, initially $(0, p)$
                     \State \T $currTag \in \mathcal{T} $ initially $\bot$
	}\EndPart \label{line:abdserver-state-end}
	
	\Statex\Statex
	
	\Part{Transitions}{ \label{line:abdserver-trans2}
	     
	     \Input{$\act{md-value-send}(t, v)_p$}
	     {
                \If{ $\neg failed$} 
               \State     $mCount \gets mCount +1$
                \State $mID \gets (p,mCount)$
                 \State let $D = \{s_{1}, \cdots, s_{f+1}\}$  - the subset of first $f+1$ servers of $S$
                  \State $send\_buff\gets$
                  \Statex \TT\TT~$\{((mID, (t, v),``full"),s):s\in D\}$
                \State $active \gets true$
                \State $currTag \gets t$
                \EndIf
                }\EndInput
                \Statex

	   \remove{
	        \Input{ $\act{md-value-send}(t, v)_{p}$} 
                  {
                    \State $mCount \gets mCount + 1$
                    \State $mID \gets (p, mCount)$
		   \For{$s~\text{in}~S_{1, f+1}$ } 
                       \State insert  $((mID, (t, v) ), s)$ to $send\_buff$
                     \EndFor
		    \State $active \gets true$
		 }\EndInput
		} 
		 \Statex
	        \Output{ $\act{md-value-send-ack}( t)_{p}$} 
	        {
	           \State  $\neg$ $failed$
                    \State  $active$
                    \State $send\_buff= \emptyset$
                    \State $t = currTag$
	        }
                  {
		    \State $active \gets false$
                      \State $currTag \gets \bot$
		 }\EndOutput

		  \Statex
		 \Output{ $\act{send}( (mID,  (t, v), ``full") )_{p,s}$}
		{
			\State $\neg failed$						
			\State  $((mID,(t, v),``full"),s) = first(send\_buff)$ 
		}
		{
			\State  $send\_buff\gets tail(send\_buff)$		
		}\EndOutput
		\Statex

		\Internal{ $\act{fail}_{s}$}
		{
		 	\State $\neg failed$
		}
		{
			\State $failed\gets true$
		}
		\EndInternal
		
	}\EndPart 
	}
	\end{multicols}
	\end{algorithmic}		
\end{algorithm*}

\begin{algorithm*}[!ht]
	\caption{$\srvIOA{\text{\mdvalueserver}}{\srvr}$ Automaton: Signature, State and Transitions at server $s \equiv s_i, 1 \leq i \leq n$} \label{ioa:md-value-server}
	\begin{algorithmic}[2]
	\begin{multicols}{3}
	{\scriptsize
	
\remove{
    	      \State {\bf Data Types:}              
  \State \T $M \subset M_{meta} \times (\mathbb{F}_q^k \cup \{ \bot\}) \times 
( \mathbb{F}_q  \cup \{ \bot\})$, 
                \State \T $M_{ID} \subseteq \mathcal{P} \times \mathbb{N}$,
                \State \T $S$ is the sorted IDs of servers in $\mathcal{S}$
%
       \Statex
}
	\Part{Signature}{ \label{line:abdserver-sig1}
		\State {\bf Input:}
		\State \T $\act{recv}((mID,  (t, v) , ``full''))_{r, s}$, 
		\Statex \TT\T $mID\in \msgSet_{ID}$, 
		$t \in \mathcal{T}$, $v \in V$,  $r \in S$
		\State \T $\act{recv}((mID,  (t, c) , ``coded''))_{r, s}$,
		\Statex  \TT\T $mID\in \msgSet_{ID}$, $t \in \mathcal{T}$, $c \in \mathbb{F}_q$,  $r \in S$
						\State {\bf Internal:}
		\State \T{ $\act{fail}_{s}$}
			\State {\bf Output:}
	       	 \State \T $\act{mds-value-deliver}(t,c)_{s}$, $t \in \mathcal{T}$, $c \in \mathbb{F}_q$		
		\State \T $\act{send}((mID,  (t, u)))_{s, r}$, $mID\in \msgSet_{ID}$,
		\Statex \TT\T $t \in \mathcal{T}$, $u \in  V \cup \mathbb{F}_q$,  $r \in S$

	}\EndPart \label{line:abdserver-sig-end}

	\Statex
	
	\Part{State}{ \label{line:abdserver-state2}
		\State $failed$, ~a Boolean initially $false$
		\State $status$, a key-value map, initially empty 
                  \State $content:  M_{ID} \rightarrow \msgSet \cup \{\bot\}$,  initially empty
                 \State $outQueue$, a queue, intially empty
	}\EndPart \label{line:abdserver-state-end2}
	
	\Statex
	
	\Part{Transitions}{ \label{line:abdserver-trans3}
	     
	     \Input{ $\act{recv}((mID, (t, v) ,``full"))_{r,s}$}
            {
         \If{  $\neg failed$} 
         
          \If{ $(status(mID) = \bot)$ } 

             \State let $D = \{s_{i+1}, \cdots, s_{f+1}\}$ be a subset of $S$ s.t. $|D| = f+1-i$
             \For{$ s' \in D$}
                  \State  append $(s',(mID, (t, v),``full"))$ 
                \Statex \TT\TT\TT\T  to $outQueue(mID)$
             \EndFor
             \For{$ s' \in S-D$}
               \State    append $(s', (t,\Phi_{s'}(v) ),``coded"))$
               \Statex \TT\TT\TT\T  to $outQueue(mID)$
                \EndFor
              \State $status(mID) \gets \act{sending}$
               \State  $content(mID) \gets (t,\Phi_s(v))$
             \EndIf
             \EndIf
          }\EndInput
       \Statex
  
          \Input{ $\act{recv}((mID, (t,c) ,``coded"))_{r,s}$}
         {
         \If{ $\neg  failed$ }
           \If{ $status(mID) \neq \act{delivered}$ }
                \State $status(mID) \gets \act{ready}$
                 \State $content(mID) \gets (t,c)$
            \EndIf
            \EndIf
	    }\EndInput
	     \Statex
	     \Output{ $\act{send}((mID, (t, u) )_{s,s'}$}
              {
                 \State $\neg failed$
                  \State $(s', (t, u)) = first(outQueue(MID))$
                  }
             {
               \State $outQueue(mID)\gets tail(outQueue(mID))$
               \If{$outQueue(mID) = \emptyset$}
                      \State     $     status(mID) \gets \act{ready}$
                \EndIf
	      }
	      \EndOutput 
	      \Statex    
	      \Output{ $\act{md-value-deliver}(t, c)_s$}
               {
                   \State $\neg failed$
                    \State $mID \in M_{ID}$
                   \State $status(mID) = \act{ready}$
                   \State $(t, c) = content(mID)$
                }
                {
                   \State $status(mID) \gets \act{delivered}$
                    \State $content(mID) \gets \bot$
	      }\EndOutput
	      \Statex		
	}\EndPart 
	}\end{multicols}
	\end{algorithmic}		
\end{algorithm*}

\subsection{\mdmetaprim~primitive} 
The \emph{\mdmetaprim}~primitive ensures that if a server $s \in S$ delivers some metadata $m$, from a metadata alphabet $M_m$,   then it is delivered at every non-faulty server $s' \in S$. 
The primitive is defined via the events $\act{md-meta-send}(m)_p$ and $\act{md-meta-deliver}(m)_p$. The difference with respect to the $\text{\mdvalueprim}$ primitive is that here we simply deliver the transmitted message $m$ itself at all the servers, while the $\text{\mdvalueprim}$  only delivered the corresponding coded-elements. Thus the implementation of \emph{\mdmetaprim}~primitive is in fact simpler; the main difference is that while sending messages to the servers in $S -D$ by a server $s_i \in D$, $s_i$ simply sends $m$, whereas in $\text{\mdvalueprim}$ protocol recall that $s_i$ calculated and sent only the corresponding coded elements.
\remove{ 
\section{$\SODA$ Algorithm }\label{sec:algo}
In this section,  we  present the  $\SODA$ algorithm.   The  algorithm  employs majority quorum, and uses  erasure codes to reduce storage cost. Detailed algorithmic steps for the reader, writer and server processes are presented in Fig. \ref{fig:soda-writer}, \ref{fig:soda-reader} and \ref{fig:soda-server}, respectively. For simplicity, we only present the pseudo-code instead of a formal description using IO Automata. SODA uses an $[n, k]$ MDS code with $k = n-f$. Atomicity and liveness are guaranteed under the assumption that at most $f$ servers crash during any execution. SODA can be designed for any $f$ such that $f \leq \frac{n-1}{2}$. 
}
\begin{algorithm}[!ht]
	\begin{algorithmic}[1]
		\begin{multicols}{2}{\footnotesize
				\Part{ \underline{\GetW}} {
					\For{$s \in {\mathcal S}$}
					\State  send(\writeGetTag) to $s$
					\EndFor
					\State  Wait to hear from a majority
					\State  Select the highest tag  $t_{max}$.
				}\EndPart
				
				\Part{\underline{\PutW}} {
					\State Create new tag $t_w = (t_{max}.z + 1, w)$.  
					\State invoke \mdvaluesend($t_w, v$)
					\State  Wait for acknowledgments from $k$ servers, and terminate
				}	\EndPart
				
			}\end{multicols}
		\end{algorithmic}	
		\caption{Protocol for   ${\mathbf{write}(v)_w}, w \in \mathcal{W}$ in  
			$\SODA$.}
		\label{fig:soda-writer}
	\end{algorithm}
	
	\begin{algorithm}[!ht]
		\begin{algorithmic}[1]
			\begin{multicols}{2}{\footnotesize			
					\Part{{\underline{{\GetR} }}}{ 
				             \For{$s \in {\mathcal S}$} 
						\State  send(\readGetTag) to $s$
						\EndFor 
						\State  Wait to hear from 
						a majority. 
						\State Select the highest tag  $t_{r}$. 
					}\EndPart
					\Statex
					
					\Part{\underline{{\GetRV} }}{
						\State invoke \mdmetasend(\readValueTag, $(r, t_r)$)
						\State  Collect  messages of form $(t, c_s)$ in set  $M = \{ (t, c_s): (t, c_s) \in \mathcal{T} \times \mathbb{F}_q \}$ 
						until there exists $M' \subseteq M$ such that   $|M'| = k$ and   $\forall m_1, m_2 \in M'$  $m_1.t = m_2.t$. 
						\State   $C \gets \bigcup_{m \in M'}\{m.c_s\}$
						\State Decode value $v \gets \Phi^{-1}(C)$.
					}\EndPart
					\Statex
					
					\Part{\underline{\CompR}} { 
						\State  invoke~\mdmetasend(\readCompleteTag, $(r, t_r)$)
						\State  return $v$. 
					}\EndPart
					
				}\end{multicols}
			\end{algorithmic}	
			\caption{The protocol for reader $\readd_r$, $r \in {\mathcal R}$ in  
				$\SODA$.}
			\label{fig:soda-reader}
		\end{algorithm}

		
		\begin{algorithm*}[!ht]
			\begin{algorithmic}[2]
				\begin{multicols}{3}{\footnotesize
						
						\Part{State Variables}{ 
							
							\Statex $(t, c_s) \in \mathcal{T} \times \mathbb{F}_q$, initially   $(t_0, c_0)$
							\Statex $R_c$,  set of pairs as $(r, t_{r})$, initially empty. 
							\Statex $H$ set of tuples $(t, s, r) $$ \in$$ 
							\mathcal{T}$$\times $$\mathcal{S}$$ \times$$ \mathcal{R}$, initially empty.
						}\EndPart
						\Statex
						\Part {\underline{On recv(\writeGetTag) from writer $w$}} {
							\Statex  Respond with locally stored tag $t$ to  writer $w$.
						}\EndPart
						\Statex
						\Part{\underline{On  $\text{md-value-deliver}(t_w,c'_s)_s$ }} { 
							\State {\bf for} $(r, t_r) \in R_c$
							\State \T{\bf if} $t_w \geq t_r$ {\bf then}
							\State \TT send $(t_w, {c'}_s)$ to  the reader $r$ 
							\State \TT $H \gets H \cup \{(t_w, s, r)\}$
							\State \TT invoke~\mdmetasend((\readDisperseTag, $(t_w,s, r)$).
							\State	{\bf if} $t_w > t$  {\bf then}  
							\State	\T  $(t, c_s) \gets (t_w, {c'}_s)$ 
							\State	 Send acknowledgment to the writer $w$.
						}\EndPart
						\Statex
						
						\Part{ \underline{recv(\readGetTag) from reader $r$ }}{
							\Statex  Respond with locally stored tag $t$ to reader $r$
						}\EndPart
						\Statex
						\Statex
						\Part{ \underline{On $\text{md-meta-deliver}(\text{\readValueTag}, (r,t_r))_s$  }} {
							\State {\bf if} $(t_0, s, r) \in H$ {\bf then} 
							\State	\T $ H_{r} \overset{def}{=}  \{ (\hat{t},\hat{s},\hat{r}) \in H : \hat{r} = r\}$ //temp variable 
							\State	\T $H \gets H \backslash H_r$ 
							\State	{\bf else} 
							\State	\T $R_c \gets R_c \cup \{ (r, t_r)\}$ 
							\State	\T {\bf if}  $t \geq t_r$ {\bf then}
							\State	\TT send $(t, c_s)$ to reader $r$
							\State	\TT $H \gets H \cup \{(t, s, r)\}$
							\State	\TT invoke~\mdmetasend((\readDisperseTag, $(t,s, r)$).
						}\EndPart	
						\Statex		
						\Part{\underline{On $\text{md-meta-deliver}(\text{\readCompleteTag}, (r, t_r))_s$}}{ 
							\State	{\bf if} $(r, t_{r}) \in R_c$ for some tag $t_{r}$ {\bf then} 
							\State	\T $ R_c \gets R_c \backslash \{ (r, t_{r})\}$ 
							\State	\T $ H_{r} \overset{def}{=} \{ (\hat{t},\hat{s},\hat{r}) \in H : \hat{r} = r\}$
							\State		\T $H \gets H \backslash H_r$ 
							\State		{\bf else} 
							\State		\T $H \gets H  \cup \{(t_0, s, r)\}$
						}\EndPart	
						\Statex	
						
						\Part { \underline{On $\text{md-meta-deliver}(\text{\readDisperseTag}, (t,s',r))_s$ } }{ 
							\State		$H \gets  H \cup \{(t, s', r)\}$ 
							\State		{\bf if} $(r, t_{r}) \in R_c$  {\bf then} 
							\State		\T $ H_{t,r} \overset{def}{=}  \{ (\hat{t},\hat{s},\hat{r}) \in H : \hat{t} = t, \hat{r} = r\}$
							\State		\T {\bf if} $|H_{t,r}| \geq k$ {\bf then} 
							\State		\TT $R_c \gets R_c \backslash \{ (r, t_{r})\}$ 
							\State		\TT $ H_{r} \overset{def}{=}  \{ (\hat{t},\hat{s},\hat{r}) \in H : \hat{r} = r\}$
							\State		\TT  $H \gets H \backslash H_r$ 
						}\EndPart		
						
					}\end{multicols}
				\end{algorithmic}	
				\caption{The protocol for server $s \in {\mathcal S}$ in  
					$\SODA$ algorithm in the MWMR setting.}\label{fig:soda-server}
			\end{algorithm*}		
		\section{$\SODA$ Algorithm }\label{sec:algo}
In this section,  we  present the  $\SODA$ algorithm.   The  algorithm  employs majority quorum, and uses  erasure codes to reduce storage cost. Detailed algorithmic steps for the reader, writer and server processes are presented in Fig. \ref{fig:soda-writer}, \ref{fig:soda-reader} and \ref{fig:soda-server}, respectively. For simplicity, we only present the pseudo-code instead of a formal description using IO Automata. SODA uses an $[n, k]$ MDS code with $k = n-f$. Atomicity and liveness are guaranteed under the assumption that at most $f$ servers crash during any execution. SODA can be designed for any $f$ such that $f \leq \frac{n-1}{2}$. 	
			For version control of the  object values  we use tags.  A tag $t$ is defined as a pair $(z, w)$, where $z \in \mathbb{N}$ and $w \in \mathcal{W}$  ID of a writer. We use $\mathcal{T}$ to denote the set of all possible tags. For any two tags $t_1, t_2 \in \mathcal{T}$ we say  $t_2 > t_1$ if $(i)$ $t_2.z > t_1.z$ or $(ii)$ $t_2.z = t_1.z$ and $t_2.w > t_1.w$.
			
			Each server stores three state variables: $(i)$ $(t, c_s)$,   tag and  coded element pair,  which is initially set to $(t_0, c_0)$,  $(ii)$ $R_c$,  a  set of pairs  of the form $(r, t_{r})$, where the pair $(r, t_{r})$ indicates the fact that the reader $r$ is being currently served by this server. Here $t_{r}$ denotes the tag requested by the reader $r$. Initially, $R_c = \emptyset$, $(iii)$ $H$, a set of tuples $(t, s', r)$ that is used to  indicate the fact that the server $s'$ has sent a coded element corresponding to the tag $t$, to reader $r$ . Initially, $H = \emptyset$. 
			
			Two types of messages are sent, messages that carry metadata, and messages that comprise  in part or full an object value. The messages sent from the clients are
			labeled with phase names, viz., {\readGetTag}, {\readValueTag}, {\readCompleteTag} and {\writeGetTag}. The server to server messages are labeled as {\readDisperseTag}. Also, in some phases of {\SODA},  the message-disperse primitives {\mdmetaprim} and {\mdvalueprim} are used  as services. 
			
			\paragraph{Write Operation} Assume that a writer $w$ wishes to write a value $v$. Recall that an $[n, k]$ MDS code  creates $n$ coded elements after encoding $v$. The goal is to store one coded element per server.  In order to optimize storage cost, at any point of the execution, each server only stores the coded element corresponding to one particular tag. 
			The write operation consists of two phases. In the first phase, the writer queries  all servers for the local tags that are stored, awaits response from a majority and then picks the highest tag $t_{max}$. The writer $w$ creates a new tag given by $t_w = (t_{max}.z + 1, w)$.  In the second phase, the writer  sends the message $(t_w, v)$ to all servers in $S$, via  $\act{md-meta-send}(t_w, v)$, and this ensures that every server that is non-faulty will eventually receive the message $(t_w, c_s)$, where $c_s = \Phi_s(v)$ denotes the coded element corresponding to server $s$. If the server $s$ finds that $t_w > t$, then the local tag and coded element are replaced by $(t_w, c_s)$. In any case, the server sends an acknowledgment back to the writer $w$. A few additional steps are performed by the server while responding to the  message $(t_w, c_s)$ (in response $3$, Fig. \ref{fig:soda-server}). These will be explained as part of the read operation.  Finally, the writer terminates after receiving acknowledgment from at least $k$ servers.
			
			\paragraph{Read Operation} Like a writer, a reader $r$ during the first phase polls all the servers for the locally stored tags, awaits response from a majority and then picks the highest tag, which we call here as $t_r$.  In the second phase, the reader sends the message $m = (r, t_r)$ to all servers in $\mathcal{S}$, via $\act{md-meta-send}(\text{\readGetTag}, (r, t_r))$. The algorithm is designed so that  $r$ decodes a value corresponding to some tag $t \geq t_r$. Any server that receives $m$ {\em registers} the $(r, t_r)$ pair locally. Here, we use the term \emph{register}  $(r, t_r)$ to mean adding the pair $(r, t_r)$ to $R_c$ by executing the 
			step $R_c \gets R_c \cup \{ (r, t_r)\}$ during the {\GetRVResp} phase at the server.
			Similarly, by \emph{unregister} we mean the opposite, i.e., remove the pair from $R_c$.
			The server sends the locally available $(t, c_s)$ pair to the reader if $t \geq t_r$. Furthermore, every time a new message $(t_w, c_s)$ is received at the server, due to some concurrent write with $(t_w, v)$ , the server sends the message $(t_w, c_s)$ to  $r$ if $t_w \geq t_r$. Note that there can be situations where the server does not store $c_s$ locally,  for instance, if the local tag $t$ is higher than the writer's tag $t_w$, but simply sends the coded element $c_s$ to  $r$. The reader keeps accumulating  $(t, c_s)$ pairs it receives from various servers, until the reader has $k$ coded elements corresponding to some tag $t_{read}$.  At this point the reader decodes the value $(t_{read}, v)$. Before returning the value $v$, the reader sends a \text{\readCompleteTag} message, by calling $\act{md-meta-send}(\text{\readCompleteTag}, (r, t_r))$,   to  the servers, so that,  the reader can be {\em unregistered} by the active servers, i.e.,  $(r, t_r)$ is removed from their local variable $R_c$.
			
			The algorithm ensures that a failed reader is not sent messages indefinitely by any server.
 Assume that the pair $(r, t_r)$ is registered at server $s$, to continue sending coded elements from new writes for tags higher than or equal to  $t_r$. Once $k$ distinct coded elements for such a tag is known to have been sent, reader  $r$ will be unregistered and server $s$ no longer sends messages for that read. In order to implement this, any server $s'$ that sends a coded element corresponding to tag $t'$ to reader $r$ also sends $(s', t', r)$ to all the other servers, by calling $\act{md-meta-send}(\text{\readDisperseTag}, (s', t', r))$. The server $s$ which receives the $(s', t', r)$ tuple adds it to a local history variable $H$, and is able to keep track of the number of coded elements sent to the registered reader $r$. So,  server $s$ eventually unregisters reader $r$ and also cleans up history variable $H$ by removing the tuples corresponding to $r$. 
			
\paragraph{Additional Notes on SODA}
$(1)$ Server $s$ accumulates any received $(s', t', r')$ tuple in its history variable $H$, even if reader $r'$ has not yet  been registered by it. The use of the {\em message-disperse} primitive by $r'$, by calling $\text{md-meta-send}(\text{\readValueTag}, (r', t_{r'}))$,  to register the pair $(r', t_{r'})$ ensures that  $s$ will also eventually register $r'$. Once $r'$ gets registered at $s$, these entries will be used by $s$ to figure out if $r'$ can be unregistered. 
			
$(2)$ Since we do not assume any order in  message arrivals,  a \text{\readCompleteTag} message may arrive at server $s$ from reader $r$ even before the server $s$ receives the request for registration from $r$. In this case,  during the response to \text{\readCompleteTag} phase, the server adds the tuple $(t_0, s, r)$ to the set variable $H$, where $t_0$ is a dummy tag. If the server is non-faulty, we  know that the registration request from the reader will arrive at $s$ at some future point in time. The reader $r$ is registered by server $s$ in  response to {\GetRVResp} phase only if the  tuple $(t_0, s, r)$ is not in  $H$.
			
$(3)$ During each read operation the reader  appends a unique identifier (eg: a counter or a time stamp) in addition to its own id $r$.  Though we show in the next Section that every server will eventually stop sending coded elements to any reader $r$,  it can happen that the entries in $H$ corresponding to $r$ are not entirely cleared. The usage of unique identifiers for distinct read operations from the same reader 
ensures that the stale entries in $H$ do not affect new reads. To keep the presentation simple,  we do not  explicitly indicate these identifiers in Fig. \ref{fig:soda-reader}.

\section{Analysis of SODA} \label{sec:analysis}
In this section, we present our claims regarding  the liveness and atomicity properties of the SODA algorithm. We also give bounds on the storage and communication costs. 
			
\subsection{Liveness and Atomicty}
Recall that by liveness, we mean that during any execution of the $\SODA$,  any read or write operation initiated by non-faulty reader or writer completes,  despite the crash failure of any other client and up to $f$ server crash failures.
\vspace{0.05in}		
\begin{theorem}  \label{thm:liveness_soda}
Let $\beta$ be a well-formed execution of $\SODA$. Also, let $\Pi$ denote the set of all client operations that take place during the execution. Then every operation $\pi \in \Pi$ associated with a non-faulty client completes.
\end{theorem}
\vspace{0.05in}			
In order to prove the atomicity property of SODA  for any well-formed execution $\beta$, we define a partial order ($\prec$) in $\Pi$  and then show  that $\prec$  satisfies the properties $P1$, $P2$ and $P3$ given in Lemma \ref{lem:atom}. For every operation $\pi$ in $\Pi$ corresponding to a non-faulty reader or writer, we associate a $(tag, value)$ pair that we denote as $(tag(\pi), value(\pi))$. For a write operation $\pi$,  we define the $(tag(\pi), value(\pi))$ pair as the message $(t_w, v)$ which the writer sends in the {\PutW} phase. If $\pi$ is a read, we define the $(tag(\pi), value(\pi))$ pair as $(t_{read}, v)$ where $v$ is the value that gets returned in the {\CompR} phase, and $t_{read}$ is the  associated tag. The partial order ($\prec$) in $\Pi$ is defined as follows: For any $\pi, \phi \in \Pi$, we say $\pi \prec \phi$  if one of the following holds: $(i)$  $tag(\pi)  < tag(\phi)$, or $(ii)$ $tag(\pi) = tag(\phi)$, and  $\pi$ and $\phi$ are write and read operations, respectively.

\vspace{0.05in}
			
\begin{theorem} \label{thm:atomicity}
Any well-formed execution  $\beta$ of the  SODA algorithm respects the atomicity properties $P1$, $P2$ and $P3$ given in Lemma \ref{lem:atom}.
			\end{theorem}

\subsection{Storage and Communication Costs}	
Below we state  the storage cost associated with {\SODA}.  Recall our assumption that, for storage cost,  we count only the data corresponding to coded elements that are locally stored, and  storage cost due to meta-data and temporary variable are ignored.

\vspace{0.05in}			

\begin{theorem} \label{thm:storage_SODA}
The worst-case total storage cost of SODA algorithm is given by $\frac{n}{n-f}$.
\end{theorem}

We next state  the communication cost for the write and read operations in  {\SODA}. Once again, note that we ignore the communication cost arising from exchange of meta-data.

\vspace{0.05in}

\begin{theorem} \label{thm:write_cost}
The communication cost of a successful write in SODA is upper bounded by $5f^2$, i.e., $O(f^2)$.
\end{theorem}

Towards deriving the read communication cost, we first observe the fact that no reader will be indefinitely sent messages by any non-faulty server.

\vspace{0.05in}

\begin{theorem} \label{thm:read_complete}
During the execution of {\SODA} algorithm, any non-faulty server which registers a reader also unregisters it, eventually.
\end{theorem}
			
\vspace{0.05in}		

\paragraph{Number of Writes Concurrent with a Read} : 
Consider a read operation initiated by a reader $r$. Let $T_{1}$ denote the earliest time instant at which the reader $r$ is registered by at least one of the servers. Also, let $T_2$ denote the earliest time instant at which $r$ is unregistered by all non-faulty servers. From Theorem \ref{thm:read_complete}, we know that the time instant $T_2$ indeed exists (i.e., it is finite). We define the parameter $\delta_w$ as the number of write operations which get initiated during the time interval $[T_1 \ \ T_2]$. The following theorem bounds the communication cost for a read operation in terms of  $\delta_w$.		

\vspace{0.05in}

\begin{theorem} \label{thm:read_cost_soda}
In SODA algorithm, the communication cost associated with a read operation is at most $\frac{n}{n-f}(\delta_w +1)$.
\end{theorem}
			
\subsection{Latency Analysis} \label{sec:partial_sync}
			
In this section, we provide conditional latency bounds for  successful read/write operations in {\SODA}. Although {\SODA} is designed for asynchronous  message passing settings, in the case of a reasonably well-behaved network we can bound  the latency of an operation. Assume that any message sent on a point-to-point channel is delivered at the corresponding destination (if non-faulty) within a duration  $\Delta > 0$, and  local computations take negligible amount of time compared to $\Delta$.  We do not assume knowledge of $\Delta$ inside the algorithm.  Thus,  latency in any  operation is  dominated by the time take taken for the delivery of all point-to-point messages involved. Under these assumptions, the  latency bounds for successful write and  read operations in {\SODA} are as follows.

\vspace{0.05in}
			
\begin{theorem} \label{thm:partial_sync_write_time}
The duration of  a successful write and read operation in {\SODA} is at most  $5\Delta$ and $6\Delta$, respectively.
\end{theorem}
			
\section{SODA$_{\text{err}}$  for Handling Errors and Erasures} \label{sec:soda_err}
In this section, we explain the usage of $[n, k]$ MDS codes for the $\text{SODA}_{\text{err}}$ algorithm. Here the parameter $k$ is chosen as $k = n - f - 2e$. The encoding and distribution of $n$ coded elements among the $n$ servers remain same as above. While decoding, we require that we are able to tolerate any $f$ missing coded elements as well as $e$ erroneous coded-elements among the remaining elements. For example, assume that $c_1, \ldots, c_{n-f}$ are available to the decoder - the servers which store the remaining coded elements might have crashed, where $e$ out of these $n-f$ elements are erroneous, and the decoder does not know the error locations. It is well known that $[n, k]$ MDS codes can tolerate any pattern of $f$ erasures and $e$ errors if $k = n - f - 2e$. We use $\Phi_{err}^{-1}$ to denote the decoder used to recover the value $v$; in this example we have $v = \Phi_{err}^{-1}(\{c_1, \ldots, c_{n-f}\})$. Once again, we make the assumption that the decoder is aware of the index set $\mathcal{I}$ corresponding to the $n-f = k+2e$ coded elements that are being used in the decoder.
			
Now we describe the modifications needed in $\SODA$  to  implement  $\text{SODA}_{\text{err}}$. In SODA, read errors can occur during the {\GetRVResp} phase, where the server is expected to send the locally stored coded element to the reader. We do not assume any error in situations where the server is only relaying a coded element, in response to the {\GetWResp} phase, since this does not involve local disk reads. Also, we assume that tags are never in error, because tags being negligible in size can be either stored entirely in memory, or replicated locally for protection against disk read errors. SODA$_{\text{err}}$ is same as SODA except for two steps (Fig. \ref{fig:sodae}), which we describe next. 

$(i)$ {\GetRV} phase initiated by the reader: Any reader must   wait until it accumulates $k + 2e$ coded elements corresponding to a  tag before it can decode. Recall that in the SODA algorithm, we only needed $k$ coded elements before the reader can decode. Also note that the decoder for the SODA$_{\text{err}}$ (which we denote as $\Phi_{err}^{-1}$) is different from that used for SODA, since now we must accept $k + 2e$ coded elements of which  certain $e$ elements are possibly erroneous. 

$(ii)$ {On  $\text{recv}(\text{\readDisperseTag} ,(t,s',r))$}:  A server checks if the number of coded elements sent (from various servers) to reader $r$ corresponding to tag $t$ is at least $k +  2e$, before deciding to unregister the reader $r$.		We now state our claims regarding the performance guarantees of the SODA$_{\text{err}}$.		
\begin{algorithm}[!ht]
				\begin{algorithmic}[1]
					\begin{multicols}{2}{\footnotesize
							
							\Part{	$\underline{{\mathbf \readd_r}}, r \in \mathcal{R}$ } {
								
								\Statex {\underline{{\GetRV} }}:
								 \State invoke~\mdmetasend((\readValueTag, $(r, t_r)$))
								\State  Collect  messages of form $(t, c_s)$ in set  $M = \{ (t, c_s): (t, c_s) \in \mathcal{T} \times \mathbb{F}_q\}$ 
								until there exists $M' \subseteq M$ such that   $|M'| = k+2e$ and   $\forall m_1, m_2 \in M'$  $m_1.t = m_2.t$.
								\State   $C \gets  \bigcup_{m \in M'}\{m.c_s\}$.
								\State Decode value  $v \gets \Phi_{err}^{-1}(C)$.
							}\EndPart

							
							\Part{$\underline{{\mathbf{\text{{\bf at server}}}_s}},  s \in \mathcal{S} $} {
								\State		{ \underline{On  $\text{recv}(\text{\readDisperseTag} ,(t,s',r))$ }}: 
								\State	$H \gets  H \cup \{(t, s', r)\}$  
								\State	{\bf if} $(r, t_{r}) \in R_c$  {\bf then} 
								\State	\T $ H_{t,r} \overset{def}{=}  \{ (\hat{t},\hat{s},\hat{r}) \in H : \hat{t} = t, \hat{r} = r\}$
								\State	\T {\bf if} $|H_{t,r}| \geq k + 2e$ {\bf then} 
								\State	\TT $R_c \gets R_c \backslash \{ (r, t_{r})\}$ 
								\State	\TT $ H_{r} \overset{def}{=}  \{ (\hat{t},\hat{s},\hat{r}) \in H : \hat{r} = r\}$ 
								\State	\TT  $H \gets H \backslash H_r$      
							}\EndPart
							
						}\end{multicols}
					\end{algorithmic}	
					\caption{The modified steps for ${\SODA}_{err}$ algorithm.}\label{fig:sodae}
				\end{algorithm}		

\begin{theorem}  {\bf (Liveness):}
					Let $\beta$ be a well-formed execution of the SODA$_{\text{err}}$ algorithm. Then every operation $\pi \in \Pi$ associated with a non-faulty client completes.
				\end{theorem}
\begin{theorem} {\bf (Atomicity):}
					Any well-formed execution fragment $\beta$ of the SODA$_{\text{err}}$  respects atomicity properties.
				\end{theorem}

\vspace{0.05in}
	
				\begin{theorem}
						$(i)$ The total storage cost of SODA$_{\text{err}}$  is  $\frac{n}{n-f-2e}$.
						$(ii)$ The write cost of SODA$_{\text{err}}$ is at most  $5f^2$, i.e, $O(f^2)$.
						and $(iii)$ The read cost of SODA$_{\text{err}}$ is  $\frac{n}{n-f-2e}(\delta_w  + 1)$, where $\delta_w$ is the number of writes which are concurrent with a read. The definition of $\delta_w$ is the same as in  SODA.
				\end{theorem}
				\section{Conclusion} \label{sec:conclusions}
				In this paper, we proposed the SODA algorithm based on $[n, k]$ MDS codes to emulate shared atomic objects in asynchronous DSSs.  SODA tolerates $f = n-k$ crash failures and achieves an optimized storage cost of $\frac{n}{n-f}$. ${\SODA}_{err}$, a modifiction of  ${\SODA}$,   which tolerates both crash failures and data read errors. Next  we plan to extend this work to $(i)$ dynamic settings where servers enter or leave the system, and $(ii)$ scenarios where background repairs are carried out to restore the contents of a crashed server.
				
				{\small
				\bibliographystyle{IEEEtran}
				\bibliography{biblio}
				}
		
			\end{document}